\documentclass[%
reprint,
 amsmath,amssymb,
 aps, physrev,
]{revtex4-2}

\usepackage{graphicx}
\usepackage{dcolumn}
\usepackage{bm}
\usepackage{xcolor}
\usepackage{soul}
\usepackage{hyperref}

\begin{document}

\preprint{APS/123-QED}

\title{\textbf{Effective transport barriers in the biquadratic nontwist map} 
}%

\author{Gabriel C. Grime}
\email{gabrielgrime@gmail.com}
\affiliation{Institute of Physics, University of São Paulo, São Paulo, Brazil}
\author{Ricardo L. Viana}
\affiliation{Physics Department, Federal University of Paraná, Curitiba, Brazil}
\affiliation{Centro Interdisciplinar de Ciência, Tecnologia e Inovação, Núcleo de Modelagem e Computação Científica, Federal University of Paraná, Curitiba, Brazil}
\author{Yves Elskens}
\affiliation{Aix-Marseille Université, UMR 7345 CNRS, PIIM, campus Saint-Jérôme case 322, 13397 Marseille cx 13, France}
\author{Iberê L. Caldas}%
\affiliation{Institute of Physics, University of São Paulo, São Paulo, Brazil}

\date{\today}

\begin{abstract}
Nontwist area-preserving maps violate the twist condition at specific orbits, resulting in shearless invariant curves that prevent chaotic transport. Plasmas and fluids with nonmonotonic equilibrium profiles may be described using nontwist systems, where even after these shearless curves breakup, effective transport barriers persist, partially reducing transport coefficients. Some nontwist systems present multiple shearless curves in phase space, increasing the complexity of transport phenomena, which have not been thoroughly investigated until now. In this work, we examine the formation of effective transport barriers in a nontwist area-preserving mapping with multiple shearless transport barriers. By quantifying the effectiveness of each transport barrier in phase space, we identify two scenarios where particular barriers dominate over others. Our results also reveal configurations where the interplay of two transport barriers creates regions in phase space with significant orbit trapping, influencing overall transport dynamics.
\end{abstract}

\maketitle


\section{Introduction}

Transport phenomena play a fundamental role in nature and involve redistributing quantities such as particles, charge, and energy. Different mechanisms are responsible for transport processes, which can explain phenomena ranging from microscopic interactions in semiconductors\,\cite{semiconductor2002} to large-scale planetary dynamics\,\cite{transpOrb2000}.

In dynamical systems, the problem of transport involves quantifying the collective motion of an ensemble of orbits between regions in phase space\,\cite{meiss2015}. Hamiltonian systems often represent models of physical significance, such as fluid advection\,\cite{morrison1998,viana2024}, \textcolor{black}{structures in solid state \cite{aubry1983}, motion of cold atoms in optical lattices \cite{lazarotto2024}} and magnetically confined plasmas\,\cite{white1984,cary2009,boozer1981,viana2023}. Featuring a mixed phase space, Hamiltonian dynamics exhibit periodic, quasiperiodic, and chaotic trajectories, with chaotic trajectories being responsible for transport\,\cite{meiss1992}.

The intermixing of regular and chaotic orbits in phase space complicates the transport problem in Hamiltonian systems. Certain structures in phase space can reduce or even eliminate chaotic transport. For instance, \textcolor{black}{considering two-dimensional maps}, quasiperiodic invariant curves act as total barriers, eliminating transport through them\,\cite{mackay1984}. Therefore, the breakup of the last invariant curve is of great importance, and in twist systems, the Kolmogorov-Arnold-Moser (KAM) theorem addresses this issue\,\cite{guckenheimer}.

Nontwist systems violate the twist condition at some orbits, forming the so-called shearless invariant curves. Although the KAM theorem is not valid in these maps, analytical and numerical results indicate that the shearless invariant curve is among the last invariant tori to break up\,\cite{rypina2007,diego1996}. Furthermore, nontwist systems have degenerate Hamiltonians, leading to new topological processes involving isochronous island chains, for example, periodic orbit collision and separatrix reconnection\,\cite{egydio1992,wurm2005}.

Nontwist dynamics appears in various research areas, including fluid advection\,\cite{diego1993,viana2024}, geophysical zonal flows\,\cite{diego2000}, and magnetically confined plasmas\,\cite{caldas2012,morrison2000}.

Even after their breakup, remnants of invariant curves, including the shearless \textcolor{black}{one}, can reduce transport coefficients in the region, forming a partial or effective transport barrier\,\cite{meiss2015}. Furthermore, the effectiveness of these barriers is closely related to manifolds crossing of isochronous islands, which can \textcolor{black}{occur} in \textcolor{black}{homoclinic or heteroclinic} topology \,\cite{corso1998,mugnaine2018,mugnaine2024,szezech2009}.


Nontwist area-preserving maps have been used to investigate the general properties of such systems. The Standard Nontwist Map, for example, is a paradigmatic system that captures the essential behavior of nontwist systems that violate the twist condition at \textcolor{black}{only one invariant curve}\,\cite{diego1996,petrisor2001,wurm2005}. Consequently, many works on effective transport barriers in nontwist systems have utilized this map\,\cite{szezech2009,szezech2012,farazmand2014,mugnaine2018,mugnaine2024}.

Recently, experimental evidence has indicated the existence of more than one transport barrier in nonmonotonic plasma equilibrium\,\cite{joffrin2003}. Additionally, plasma-transport models have \textcolor{black}{used} nontwist systems to explain transport reduction\,\cite{grimeJPP,diego2012,fonseca2014}. In such nontwist systems, more than one orbit violates the twist condition, leading to complex nontwist processes with unique characteristics, such as reconnection-collision sequences\,\cite{howard1995,wurm2005}. Recently, the Biquadratic Nontwist Map has been used to study bifurcation processes and shearless curve \textcolor{black}{breakup} in systems with multiple shearless curves\,\cite{grime2023BNM1,grime2023BNM2}. However, there has been no study so far on how multiple effective barriers influence transport in phase space.

In this work, we investigate the formation of an effective transport barrier in the Biquadratic Nontwist Map, a prototype system with multiple shearless transport barriers. Using computational and theoretical tools, we quantify the effectiveness of each barrier individually. Our results revealed scenarios where a specific barrier dominates over the others. Furthermore, the presence of two transport barriers can create chaotic regions where orbits remain trapped for extended periods.

The rest of the paper is organized as follows. Section \ref{sec_bnm} presents the area-preserving map used in our analysis. The theoretical background about transport analysis tools and quantifiers is provided in Section \ref{sec_framework}. Section \ref{sec_scenarios} applies these quantifiers to the Biquadratic Nontwist Map, exploring how multiple transport barriers affect low and high transport configurations. Finally, Section \ref{sec_conclusion} offers our conclusions.

\section{Multiple shearless curve systems}
\label{sec_bnm}

Let us consider a two-dimensional area-preserving map with a particular functional form, defined by the recurrence relations
\begin{subequations}
\label{eq_sympMap}
\begin{align}
x_{n+1} &= x_n + \omega(y_{n+1})\pmod{1}\\[0.1cm]
y_{n+1} &= y_n - f(x_n),
\end{align}
\end{subequations}
where $x\in [0,1)$ and $y\in \mathbb{R}$ are a pair of canonical coordinate and momentum. Its phase space is the infinite cylinder $\mathbb{S}^1\times \mathbb{R}$. Functions $f$ and $\omega$ must be sufficiently differentiable. Additionally, for such a system to be used as a model for studying Hamiltonian dynamics, we require $f$ to be a period-1 function with zero average\,\cite{meiss1992}.

The twist function $\omega$ gives the frequency of the orbits in phase space when the system is integrable, i.e., when $f(x)\equiv 0$. If $\omega$ has no extreme point, the map \eqref{eq_sympMap} is called a twist map, and satisfies the condition
\begin{equation}\label{eq_twistCond}
    \left|\dfrac{\partial x_{n+1}}{\partial y_n}\right|=\left| \omega'(y_{n+1})\right| > 0
\end{equation}
for every point in phase space\,\cite{reichl2004}.

Maps that do not satisfy the twist condition are called nontwist maps. Consequently, important results, such as the Kolmogorov-Arnold-Moser (KAM) theorem, are not valid\,\cite{lichtenberg}. Significant nontwist systems, such as the Standard Nontwist Map, violate the twist condition at only one \textcolor{black}{value $\omega (y^*)$, i.e. at one value $y_{n+1} = y^*$ corresponding to a curve $y_n = y^*+f(x_n)$, also called \emph{nonmonotone set}}\,\cite{diego1996,petrisor2001}. However, general maps might violate the twist condition at \textcolor{black}{several nonmonotone sets}, thereby increasing the complexity of nontwist phenomena presented, such as reconnection of separatrices\,\cite{howard1984}.

\subsection{The Biquadratic Nontwist Map}

It is suitable to use specific functional forms of $\omega$ and $f$ for the resultant map to possess useful properties. By choosing the twist function $\omega(y)=a(1-y^2)(1-\epsilon y^2)$ and the perturbation $f(x)=b\sin{(2\pi x)}$, we obtain the Biquadratic Nontwist Map (BNM)\,\cite{grime2023BNM1}
\begin{subequations}
\label{eq_bnmDef}
\begin{align}
x_{n+1} &= x_n + a(1-y_{n+1}^2)(1 - \epsilon y_{n+1}^2)\pmod{1}\\[0.1cm]
y_{n+1} &= y_n - b \sin{(2\pi x_n)}.
\end{align}
\end{subequations}
\textcolor{black}{It is a three-parameter family of area-preserving maps, with the range of interest defined as $a \in [0,1)$, $\epsilon \in \mathbb{R}^+$, and $b \in \mathbb{R}^+$. Dynamics outside this range are not relevant to this work, as the system either exhibits the same behavior or features only a single shearless curve. The parameters $a$ and $\epsilon$ shape the twist function, thereby altering the frequency profile of orbits in phase space. These parameters play a key role in controlling the positions of main resonances and nontwist phenomena, such as separatrix reconnection and shearless curve breakup. The amplitude of the perturbation is governed by the parameter $b$, which is responsible for the amount of chaos in the systems. Further details on the effects of each parameter on reconnections and shearless curve breakup can be found in Refs.~\cite{grime2023BNM1,grime2023BNM2}.}

\begin{figure}[htb]
    \centering
    \includegraphics[scale=1]{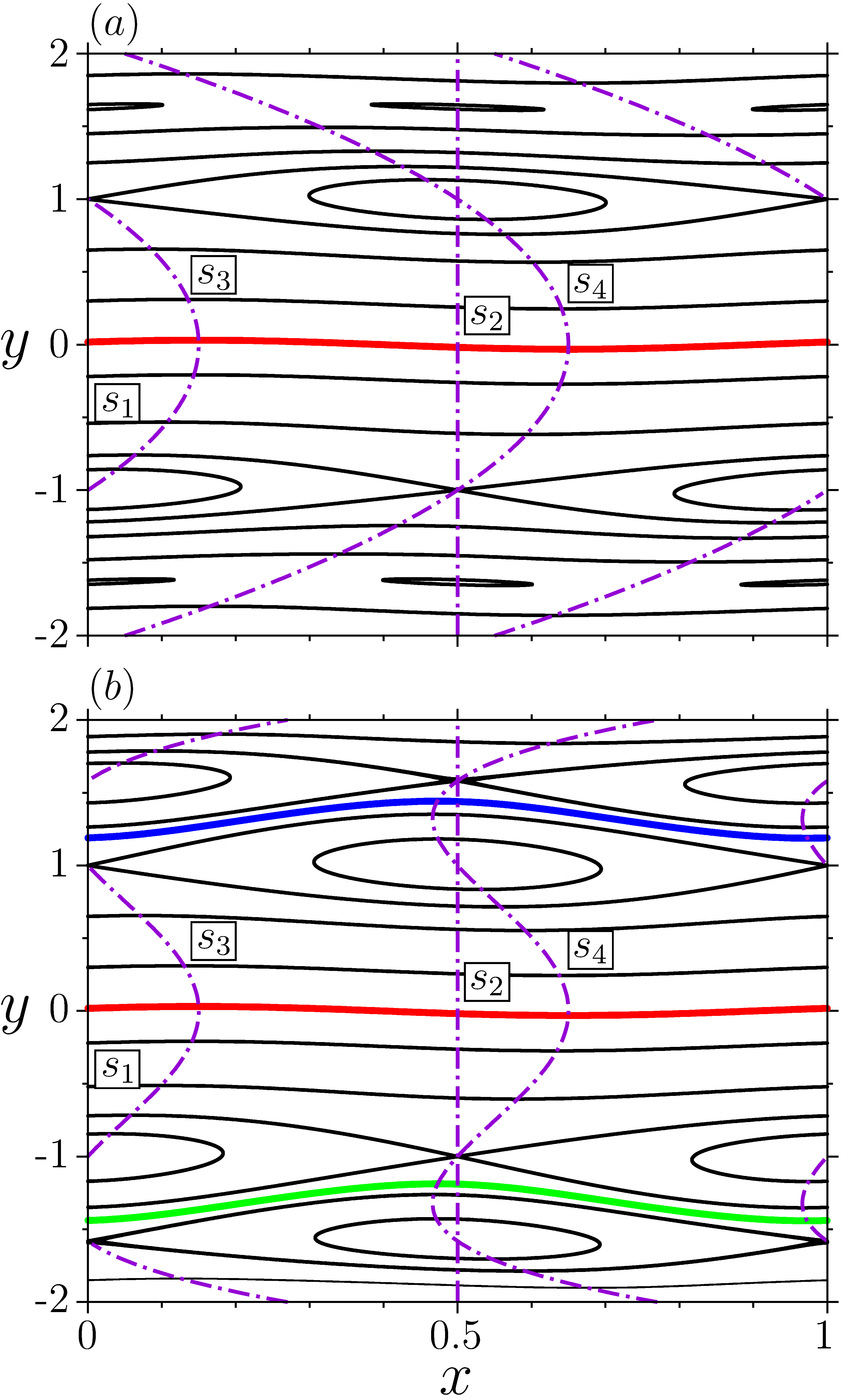}
    \caption{Phase spaces of the Biquadratic Nontwist Map, with $a=0.3$ and $b=0.05$, (a) $\epsilon=0$ and (b) $\epsilon=0.4$. Symmetry lines are marked by dashed-dotted lines, while shearless curves appear in red, blue, and green lines.}
    \label{fig_integ_PhaseSpace}
\end{figure}

When $b=0$, \textcolor{black}{the system is integrable and} the phase space contains only periodic and quasiperiodic ($y=\mathrm{constant}$) orbits. Near-integrability occurs for small perturbation parameters ($b\ll 1$). Typical phase spaces of the BNM in this regime are shown in Fig.~\ref{fig_integ_PhaseSpace}. In this case, periodic orbits give rise to resonance islands, and the quasiperiodic invariant curves become distorted.

The BNM is a nontwist map because its twist function violates the twist condition, Eq.~\eqref{eq_twistCond}. For $\epsilon>0$, the map exhibits three such orbits, known as shearless invariant curves. Each of these shearless curves intersects one of the \emph{nonmonotone sets}
\begin{subequations}\label{eq_shearlCurve}
    \begin{align}
        \mathcal{R}_0 &: y = b\sin{(2\pi x)} \\
        \mathcal{R}_{\pm} &: y = \pm \sqrt{\dfrac{1+\epsilon}{2\epsilon}} + b\sin{(2\pi x)},
    \end{align}
\end{subequations}
defined by the regions in phase space that violate the twist condition\,\cite{petrisor2001}. In this paper, we will call $\mathcal{C}_0$ the central shearless curve, associated with $\mathcal{R}_0$. The same occurs to $\mathcal{C}_\pm$, named external shearless curves, intersecting $\mathcal{R}_\pm$.

In Fig.~\ref{fig_integ_PhaseSpace}(b), the red curve stands for $\mathcal{C}_0$, while $\mathcal{C}_\pm$ are marked in blue and green. For $\epsilon=0$, Fig.~\ref{fig_integ_PhaseSpace}(a), the BNM twist function is parabolic and, in this case, the Biquadratic Nontwist Map reduces to the Standard Nontwist Map\,\cite{diego1996} (SNM), which has only a central shearless curve.

Notice that the BNM has symmetry properties concerning time evolution and spatial transformation. The time evolution symmetry leads to the symmetry lines, useful to find periodic orbits\,\cite{fuchss2006}. \textcolor{black}{Let $D=\mathbb{S}^1\times \mathbb{R}$ be the domain of the map, the symmetry lines of the Biquadratic Nontwist Map are the four sets
\begin{subequations}
\begin{align}
    s_1 &= \big\{(x,y)\in D: x=0\big\}\\
    s_2 &= \big\{(x,y)\in D: x=1/2\big\}\\
    s_3 &= \big\{(x,y)\in D: x=a(1-y^2)(1-\epsilon y^2)/2\big\}\\
    s_4 &= \big\{(x,y)\in D: x=a(1-y^2)(1-\epsilon y^2)/2 + 1/2\big\}.
\end{align}
\end{subequations}}
For example, the intersections of symmetry lines correspond to the fixed points of the BNM. Those lines are marked by dashed-dotted lines in Fig.~\ref{fig_integ_PhaseSpace}. \textcolor{black}{Furthermore, the map has the spatial symmetry
\begin{equation}\label{eq_symmetry}
    S(x,y)=(x+1/2,-y),
\end{equation}
meaning that every orbit in phase space has an associated symmetrical orbit with identical properties. For instance, the external shearless curves are symmetric, as well as the period-1 resonance islands. The only exception is the central shearless curve $\mathcal{C}_0$, which is invariant under $S$, i.e., $S(\mathcal{C}_0)=\mathcal{C}_0$, therefore it is its own symmetric~\cite{petrisor2001}. Such a property ensures that the central shearless curve possesses unique characteristics that are not shared with the external shearless curves, as will be discussed later.}

The BNM has been used in previous studies concerning nontwist systems with multiple shearless curves. Due to its symmetry properties and the range of phenomena displayed, it serves as a useful model for studying general nontwist systems, that present multiple shearless curves. Further characterization and results about the BNM can be found in Ref.~\onlinecite{grime2023BNM1,grime2023BNM2}.

\subsection{Effective transport barriers}

\begin{figure}[htb]
    \centering
    \includegraphics[scale=0.9]{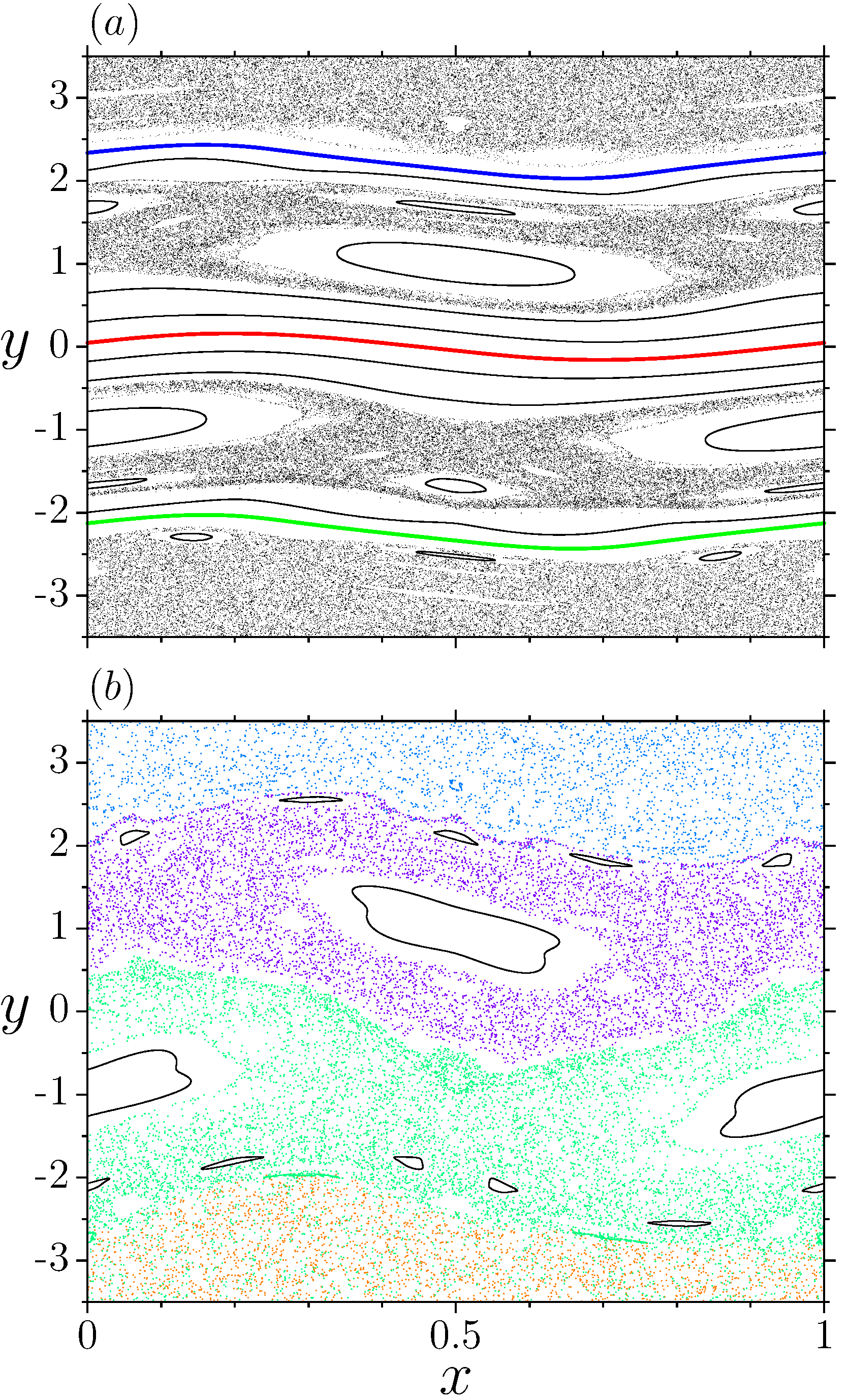}
    \caption{Phase spaces of the Biquadratic Nontwist Map for $\epsilon=0.11$, (a) with $a=0.4$ and $b=0.3$, and (b) $a=0.25$ and $b=0.7906$. A partial barrier persists once a shearless curve is broken, preventing transport between the four regions of phase space, marked by chaotic orbits of different colors.}
    \label{fig_broken_PhaseSpace}
\end{figure}

\textcolor{black}{At the range of moderate values of perturbation parameter, $b\sim 1$, the BNM exhibits a mixed-type phase space}. Alongside regular orbits (resonances and invariant curves), irregular (chaotic) orbits fill nonzero \textcolor{black}{area} regions in phase space, as depicted in Fig.~\ref{fig_broken_PhaseSpace}(a). These irregular orbits lead to chaotic transport, i.e., the motion of a collection of trajectories across different regions of phase space. As the perturbation parameter grows, invariant curves are broken, causing chaos to spread throughout phase space.  The remaining invariant curves serve as barriers to transport, delineating boundaries for chaotic orbits. Consequently, once the last rotational invariant circle is broken, chaotic orbits traverse all available space, \textcolor{black}{which excludes the islands}, leading to a scenario known as global transport.

Invariant curves are total transport barriers since they completely prevent transport along the momentum variable. However, even after their breakup, transport in the region is not diffusive. The remnants of the last invariant curve lead to a reduction in transport coefficients at the region. Such a reduction is attributed to long-time correlation functions, a signature of chaotic orbits wandering along the transport barrier in a phenomenon called stickiness\,\cite{chirikov1999}. These remnants and their influence on transport have been studied on twist\,\cite{mackay1984,meiss2015} and nontwist systems\,\cite{szezech2009,szezech2012,farazmand2014,mugnaine2018}.

Concerning nontwist systems, both analytical and numerical evidence suggest that the shearless curve is one of the last invariant curves to be broken\,\cite{rypina2007,diego1996}. In addition, the arrangement of island chains around the shearless transport barrier plays a crucial role in the effectiveness of these partial barriers\,\cite{diego2000,szezech2009}.

The Biquadratic Nontwist Map has three shearless curves that break up in different configurations, with the central and external transport barriers breaking up independently\,\cite{grime2023BNM2}. Still, there are parameter sets where all shearless curves are broken.

In the BNM, orbits can mix between four distinct regions in phase space. Figure~\ref{fig_broken_PhaseSpace}(b) illustrates such a situation, where we evolved a unique orbit in each region of phase space. These four orbits are colored as follows: (i) above the top shearless transport barrier (blue), (ii) between the top and central barriers (purple), (iii) between the central and lower barriers (green), and (iv) below the lower barrier (orange). Initially, these orbits are trapped between the barriers. However, they eventually cross the barrier, causing mixing between regions of different colors.

\section{Transport analysis framework\label{sec_framework}}

This section describes the methods used in this paper to investigate transport in the Biquadratic Nontwist Map (BNM). Some dynamical quantifiers have been used to evaluate transport properties in Hamiltonian systems. In this paper, we adopt the transmissivity of a transport barrier, escape time \textcolor{black}{of orbits} and manifold analysis.

\subsection{Transmissivity\label{sec_frameworkTransm}}

The transmissivity measures the effectiveness of a given transport barrier in preventing orbits from crossing a given region in phase space. In other words, it quantifies the strength of a transport barrier. Given a set of initial conditions, \textcolor{black}{and a maximum number of iterations $N$,} we define the transmissivity of a barrier as the fraction of orbits that cross the same barrier \textcolor{black}{after $N$ iteration}. To numerically obtain this fraction, we define the circles
\begin{equation}\label{eq_boundDef}
    \partial \mathcal{B}_\pm=\{ (x,y) \in \mathbb{S}^1\times \mathbb{R}\,:\, 0\le x < 1,\, y=\pm y_\mathrm{B}\},    
\end{equation}
in phase space, where $y_\mathrm{B}$ is a constant value. Computationally, we randomly choose a large number of initial conditions on the circle $\partial \mathcal{B}_-$, which are iterated $N$ times. The fraction of orbits that reach the circle $\partial \mathcal{B}_+$ \textcolor{black}{after, at most, $N$ iterations} is assigned as the transmissivity of the partial transport barrier within the region bounded by $\partial \mathcal{B}_\pm$. Therefore, the value of $y_\mathrm{B}$ determines which barriers of the BNM are considered, as discussed in the next section.

\textcolor{black}{Note that $\partial \mathcal{B}_\pm$ are symmetric under the transformation \eqref{eq_symmetry}, such that, $S(\partial \mathcal{B}_\pm)=\partial \mathcal{B}_\mp$. Therefore, due to the spatial symmetry of the map, the transmissivity is identical for upward and downward fluxes in phase space considering these boundaries. However, in nonintegrable Hamiltonian systems lacking spatial symmetry, the flux across a transport barrier \textcolor{black}{may exhibit} a preferred direction. Such behavior was reported in an asymmetrical Hamiltonian system under the name ratchet currents~\cite{gong2004}. In the context of nontwist systems, ratchet currents have been observed when the isochronous island chains are asymmetric~\cite{mugnaine2020}.}

A total transport barrier, which completely prevents the transport of orbits through it, has zero transmissivity \textcolor{black}{independently of the chosen maximum number of iteration $N$}. Shearless curves are examples of total transport barriers. Transmissivity values marginally greater than zero indicate a strong transport barrier, while values closer to one indicate a weak capability of preventing transport.

\begin{figure}[htb]
    \centering
    \includegraphics[scale=1]{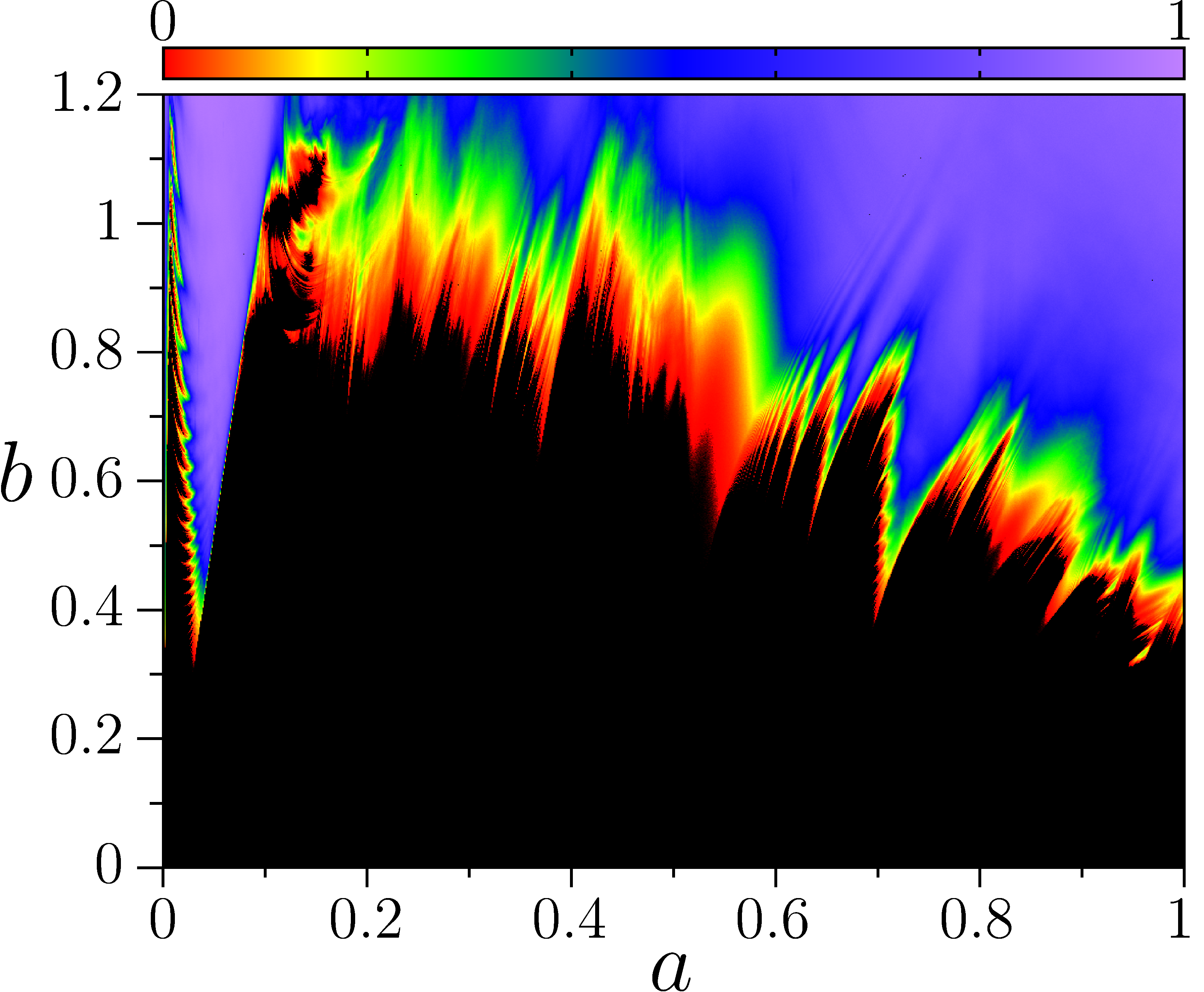}
    \caption{Transmissivity parameter space of the Biquadratic Nontwist Map, fixed $\epsilon=0.11$. Colors represent the barrier transmissivity, with black being the zero transmissivity.}
    \label{fig_transm_ParamSpace}
\end{figure}

Applying transmissivity for the BNM requires a careful choice of $y_\mathrm{B}$, as this determines which transport barrier is considered. Results concerning transmissivity in the BNM are presented in Fig.~\ref{fig_transm_ParamSpace}, \ref{fig_transm_central} and \ref{fig_transm_ext}.

\subsection{Escape time\label{sec_frameworkEscTime}}

The \textcolor{black}{escape time of each initial condition in phase space} can be used to investigate the stickiness of orbits on partial transport barriers. While transmissivity provides relevant information, it does not offer any data on the characteristic time scales associated with the transport barrier. Therefore, determining the time required for an orbit to escape a certain region of phase space \textcolor{black}{enables} us to verify the time that orbits spend in each region of phase space. This tool helps identify regions of stickiness and escape channels through which orbits leave the transport barrier.

In this work, the \textcolor{black}{escape time of initial conditions} is obtained by setting a regularly spaced grid of $2000\times 2000$ initial conditions, which are iterated a maximum of $2\cdot 10^6$ times. We compute the number of iterations needed for the corresponding orbits to exit a certain region of phase space $\mathcal{B}$, bounded by $\partial \mathcal{B}_\pm$, previously defined. The choice of the constant $y_\mathrm{B}$, which defines the region boundary, determines the transport barrier being considered. Results of the escape time are shown in Figures~\ref{fig_escTime_cent}, \ref{fig_escTime_cent_zoom}, \ref{fig_escTime_ext} and \ref{fig_escTime_ext_zoom}.

\subsection{Stable and unstable manifolds\label{sec_frameworkManif}}

Here, we provide a brief introduction to invariant manifolds in dynamical systems. Since this work focuses on the Biquadratic Nontwist Map, we restrict our discussion to two-dimensional area-preserving maps.

Let $P$ be a hyperbolic period-$p$ orbit of a two-dimensional area-preserving map $M:\,z\mapsto M(z)$, whose inverse is $M^{-1}$. \textcolor{black}{Briefly, the stable manifold $W_{\mathrm{s}}^P$ and the unstable manifold $W_{\mathrm{u}}^P$ associated with the hyperbolic orbit $P$ are \cite{guckenheimer}}
\begin{subequations}
    \begin{align}
        W_\mathrm{s}^P &= \big\{ z\in D: M^n(z)\to P,\, n\to \infty\big\}\\
        W_\mathrm{u}^P &= \big\{ z\in D: M^{-n}(z)\to P,\, n\to \infty\big\},
    \end{align}
\end{subequations}
where $D=\mathbb{S}^1\times \mathbb{R}$ is the domain of the map. In the context of our study, $W_\mathrm{s,u}^P$ are one-dimensional \textcolor{black}{sets comprising $2p$ branches}. The computational method to obtain such invariant sets starts by choosing an appropriate linear segment, whose direction is given by the eigenvectors of the associated hyperbolic orbit. This segment is then evolved under the map dynamics to obtain the unstable manifold and under its inverse to obtain the stable manifold\,\cite{ciro2017}.

Locally, the manifolds of a map give the direction of the tangent space, indicating the direction in which \textcolor{black}{nearby} orbits evolve. Furthermore, the configuration of the stable and unstable manifolds of hyperbolic orbits determines the behavior of chaotic orbits and, consequently, transport in phase space.

\section{Scenarios of dominant transport barriers}
\label{sec_scenarios}
This section examines how multiple transport barriers in the Biquadratic Nontwist Map (BNM) influence transport in phase space. Using the techniques described in the previous section, we explore how the central and external transport barriers impact low and high transport scenarios on the map.

Since the BNM features three independent shearless transport barriers, we can study the effect of each barrier individually or their combined effect. Regarding the last case, the boundaries $\partial \mathcal{B}_\pm$ (defined in Eq.~\eqref{eq_boundDef}) must extend beyond the external shearless transport barriers. Conversely, to isolate the effect of the central transport barrier, the same boundaries must be placed between the central and external barriers.

\textcolor{black}{Since the BNM has three parameters, being $a$ and $b$ directly related to the Standard Nontwist Map, we fix $\epsilon=0.11$ to enable a direct comparison of results between the two maps. For $\epsilon\ll1$, shearless curves $\mathcal{C}_\pm$ are located away from $\mathcal{C}_0$ and the \textcolor{black}{dynamics} closely resembles the Standard Nontwist Map. Conversely, for \textcolor{black}{$\epsilon\gtrsim 1$} all the shearless curves are brought very close together, potentially leading to interactions. With the chosen $\epsilon$ value, the central and external barriers are close enough to influence the dynamics but their effect on transport can be easily distinguished}.

Details about the typical phase space of the BNM in this configuration are shown in Figures~\ref{fig_broken_PhaseSpace} and \ref{fig_escTime_cent}. The external transport barriers, defined by Eq.~\eqref{eq_shearlCurve}, are located within the region $|y|\lesssim 3.35$. Therefore, choosing $y_\mathrm{B}=5$ ensures the boundaries $\partial \mathcal{B}_\pm$ are beyond the external shearless barriers. In contrast, to focus on the transmissivity of the central barrier alone, we set $y_\mathrm{B}=1.5$, positioning the boundaries between the central and external shearless barriers.

Taking into account the effect of all three shearless transport barriers, we illustrate the dependence of transmissivity on the parameters of the BNM in Figure~\ref{fig_transm_ParamSpace}. We computed the transmissivity using the method outlined in section~\ref{sec_frameworkTransm}, with a total of $10^4$ initial conditions, iterated $10^4$ times, considering boundaries $\partial \mathcal{B}_\pm$ where $y_\mathrm{B}=5$.

Black regions in parameter space have zero transmissivity, i.e., the phase space has at least one invariant curve acting as a total transport barrier. \textcolor{black}{Since we only iterated up to $10^4$ times, some numerical imprecision occurs, mostly at the boundary of black regions. More precise methods can be used to verify the existence of total transport barriers, but the results are roughly the same. More details on these methods and the scenarios with total transport barriers in the BNM can be found in Ref.~\onlinecite{grime2023BNM2}}.

According to the transmissivity parameter space, Figure~\ref{fig_transm_ParamSpace}, transport is still reduced after all invariant curves have broken. Regions with zero transmissivity are surrounded by low-transport zones, indicating that transport remains low immediately after the shearless curve breakup. Additionally, there is a noticeable sensitivity of transmissivity to the map parameters, which can vary gradually or abruptly depending on the region of the parameter space.

\textcolor{black}{Certainly the calculated transmissivity in Figure~\ref{fig_transm_ParamSpace} is dependent on the choice of maximum iteration number $N$. By changing $N$, we offset the transmissivity value and modify the gradient of the transmissivity in the parameter space. However, provided a sufficiently high $N$, these modifications will not compromise the identification of low and high transport configurations mentioned before. Such a value of $N$ depends on the escape time distribution in initial conditions, as we will see next.}

Abrupt changes in transmissivity are attributed to topological modifications in the remnants of the transport barriers\,\cite{corso1998}. Furthermore, in the BNM, those changes can be associated with modification of the central or external transport barrier. The first scenario is named the \textit{centrally dominant} transport barrier, while the last is the \textit{externally dominant} transport barrier. The characterization of the topological changes in both dominant scenarios is detailed below.

\subsection{Centrally dominant}

\begin{figure}[htb]
    \centering
    \includegraphics{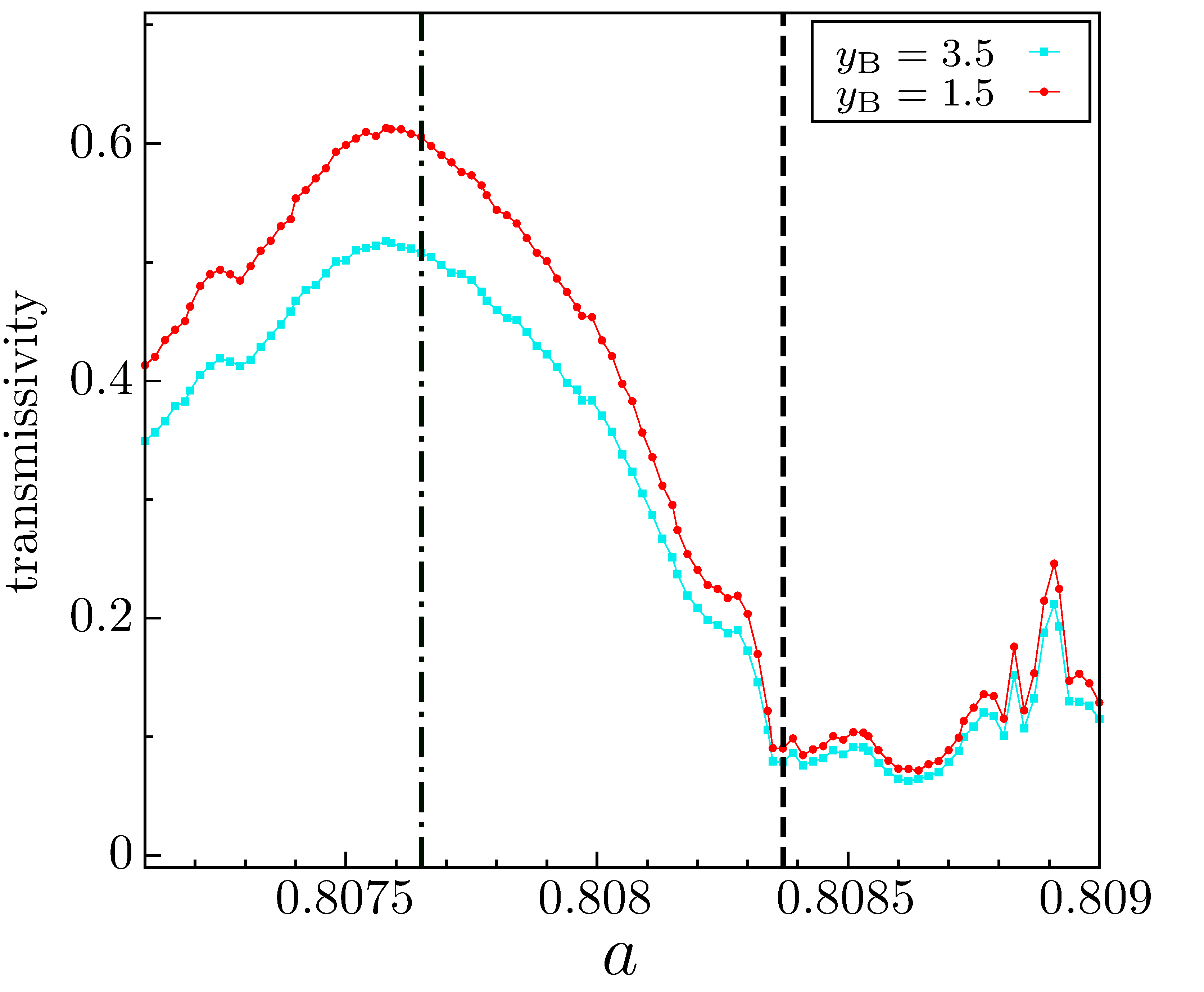}
    \caption{Transmissivity of the Biquadratic Nontwist Map, as function of $a$, fixed $b=0.58$ and $\epsilon=0.11$. Dashed-doted and dashed lines mark high and low transport configurations, respectively.}
    \label{fig_transm_central}
\end{figure}

In the centrally dominant scenario, nontwist processes involving the central transport barriers dictate the effectiveness of transport in the BNM. We compare the transmissivity when considering only the central barrier ($y_\mathrm{B}=1.5$) versus considering all the barriers ($y_\mathrm{B}=3.5$). Figure~\ref{fig_transm_central} displays the transmissivity as a function of $a$, with fixed values of $b$ and $\epsilon$. Here, the transmissivity is obtained using an ensemble of $10^5$ initial conditions iterated $10^6$ times, or until they reach the boundary at $y= y_\mathrm{B}$.

We observe an abrupt change in transmissivity for both values of $y_\mathrm{B}$. Two specific values of $a$ are highlighted: the dashed line marks a low transmissivity configuration, while the dashed-dotted line marks a high transmissivity one.

\textcolor{black}{We intentionally selected this set of parameters to demonstrate that slight changes in parameter values can significantly affect the transmissivity of the transport barriers. While other configurations with low and high transmissivity exist, showing larger or smaller transmissivity gradients, the fundamental conclusions derived from this example remain unchanged. For details on transmissivity variations over a broader parameter range, see Figure~\ref{fig_transm_ParamSpace} or consult Figure 7 in Ref.~\cite{grime2023BNM2}.}

The results indicate that both low and high transport regimes are evident for the two values of $y_\mathrm{B}$. Additionally, the transmissivity considering all transport barriers is slightly smaller compared to the effect of the central barrier alone. Briefly, in the centrally dominant scenario, transmissivity is primarily influenced by the central transport barrier. The external barriers tend to reduce transport, but their effect is minimal, especially in regions of low transmissivity.

\begin{figure}[htb]
    \centering
    \includegraphics[scale=1]{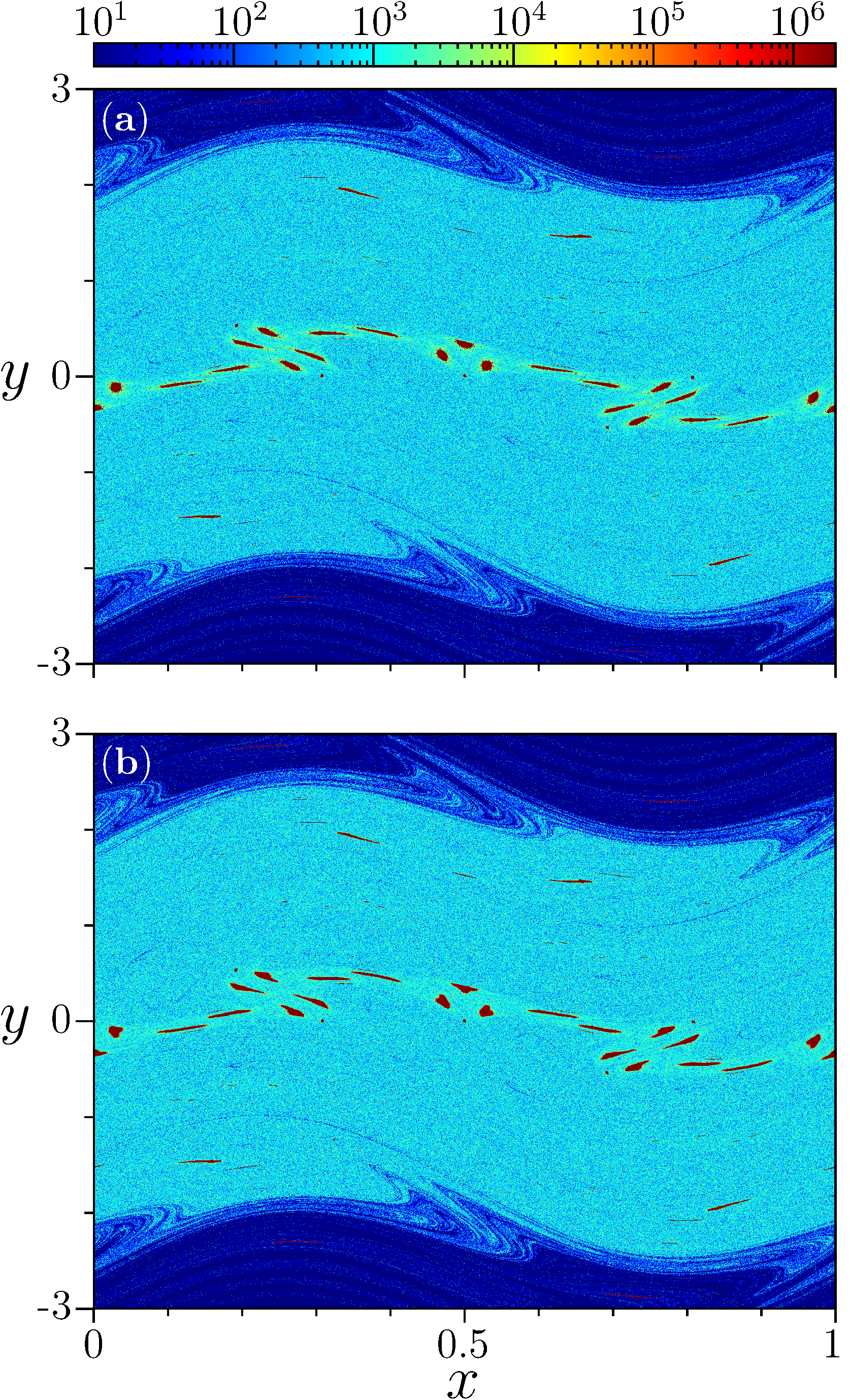}
    \caption{Escape time of the trajectories in phase space of the Biquadratic Nontwist Map, with $b=0.58$, $\epsilon=0.11$, (a) $a=0.80765$ and (b) $a=0.80837$, corresponding to high and low transport configurations of Fig.~\ref{fig_transm_central}. Here we considered $y_\mathrm{B}=3.5$.}
    \label{fig_escTime_cent}
\end{figure}

Figure~\ref{fig_escTime_cent} shows the number of iterations needed for an orbit to escape the region of phase space bounded by $\partial \mathcal{B}_\pm$, where $y_\mathrm{B}=3.5$. The parameters used correspond to high [Fig.~\ref{fig_escTime_cent}(a)] and low [Fig.~\ref{fig_escTime_cent}(b)] transmissivity configurations from Figure~\ref{fig_transm_central}. In both configurations, all invariant curves were destroyed, resulting in all chaotic orbits eventually escaping. However, the required time for these escapes greatly varies.

In both high and low transmissivity scenarios, most orbits escape after $10^3$ iterations, particularly in the region bounded by $y\approx \pm 2$. Outside this portion of phase space, orbits typically take around $10$ iterations to escape. The region where the escape time changes abruptly delineates the external partial transport barriers. Despite having different transmissivity values, Figures~\ref{fig_escTime_cent}(a) and \ref{fig_escTime_cent}(b) do not show significant differences in escape times.

\textcolor{black}{Additionally, for these two representative parameter values, the majority of orbits cross the boundaries $\delta \mathcal{B}_\pm$ within $10^4$ iterations. This indicates that, considering the transmissivity, $N=10^4$ is sufficiently large to distinguish between high and low transport configurations in the transmissivity parameter space.}

\begin{figure}[htb]
    \centering
    \includegraphics[scale=1]{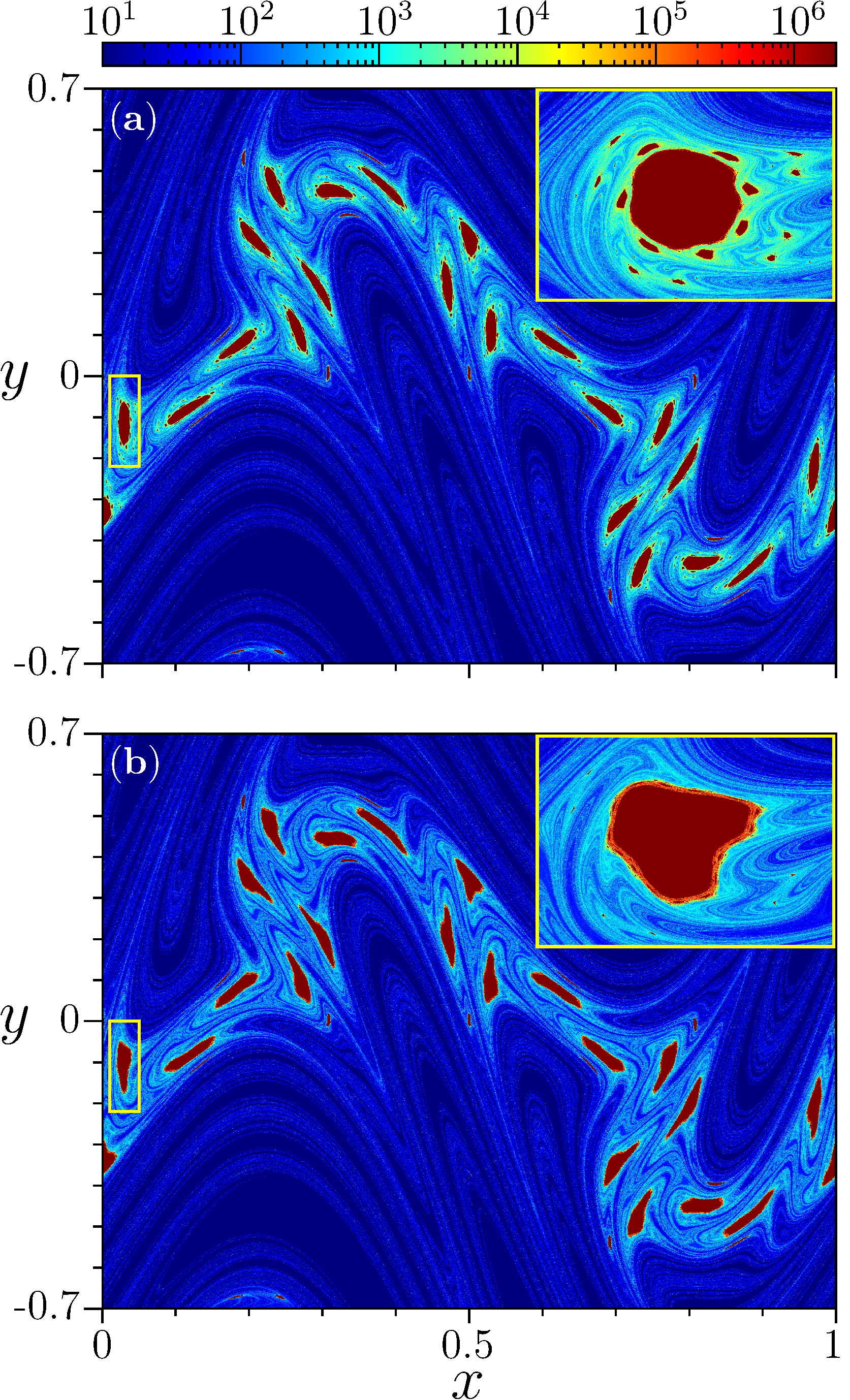}
    \caption{Escape time near the central transport barrier of the Biquadratic Nontwist Map for (a) high and (b) low transmissivity regimes of Fig.~\ref{fig_transm_ext}, considering $y_\mathrm{B}=1.5$. Magnifications of the highlighted regions are embedded in the corresponding panels.}
    \label{fig_escTime_cent_zoom}
\end{figure}

Computing the escape time to $y_\mathrm{B}=1.5$, which lies between the central and external barriers, Figure~\ref{fig_escTime_cent_zoom} shows a phase space with a considerably different escape time distribution. \textcolor{black}{The central transport barrier is formed by the manifolds associated with a pair of period-11 twin island chains embedded in the chaotic sea}. Points inside these islands do not escape since they correspond to invariant sets. However, examining the escape times near the islands (highlighted rectangles in Figure~\ref{fig_escTime_cent_zoom}), we observe that orbits adjacent to them linger longer than the rest of the chaotic orbits.

In some way, these adjacent orbits resemble the periodic behavior of the islands, causing them to remain trapped in the region for extended periods. This dynamical trap of orbits, called stickiness, has been studied in both twist\,\cite{meiss2015} and nontwist systems\,\cite{szezech2009}.

A detailed examination of Figures~\ref{fig_escTime_cent_zoom}(a) and \ref{fig_escTime_cent_zoom}(b), which correspond to high and low transmissivity \textcolor{black}{cases}, indicates distinct escape times \textcolor{black}{near} the islands. In the low transmissivity scenario, sticky orbits require approximately $10^5$ iterations to escape the central region, whereas in the high transmissivity \textcolor{black}{case}, they take only about $10^4$ iterations. Notably, for the same parameters, the escape times considering the external transport barriers remain roughly the same (see Fig.~\ref{fig_escTime_cent}). Therefore, the escape time analysis also indicates the dominance of the central barrier over the external ones in this scenario.

\begin{figure*}[t]
    \centering
    \includegraphics[scale=1]{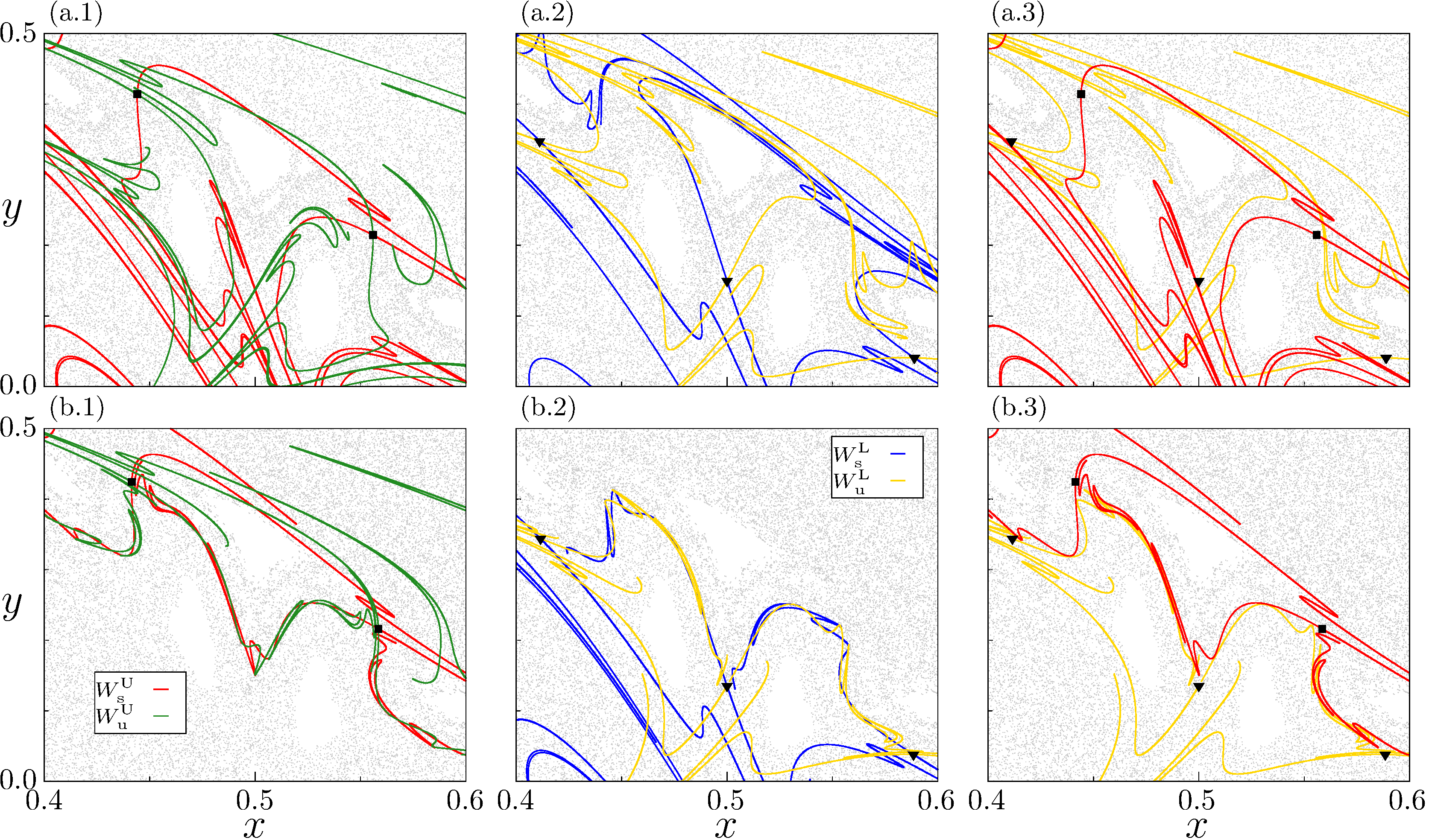}
    \caption{Stable and unstable manifolds of the upper ($W_\mathrm{s}^U$ and $W_\mathrm{u}^U$) and lower ($W_\mathrm{s}^L$ and $W_\mathrm{u}^L$) periodic orbit, considering (a) high and (b) low transport configurations of Fig.~\ref{fig_transm_central}. Chaotic orbits near them are plotted in light-grey.}
    \label{fig_manif_cent}
\end{figure*}

As defined in section~\ref{sec_framework}, manifolds dictate the behavior of orbits in phase space. According to the Poincaré-Birkhoff theorem, island chains originate from a pair of stable and unstable periodic orbits\,\cite{lichtenberg}. In turn, each unstable periodic orbit has a stable and an unstable hyperbolic manifold. Specifically, the island chains in the central transport barrier are denoted as the upper and lower chains. The hyperbolic period-11 orbit of the upper (lower) chain is marked by filled squares (triangles) and denoted by $U$ ($L$) in Figure~\ref{fig_manif_cent}. The corresponding stable and unstable manifolds are denoted by $W_\mathrm{s}^U$ ($W_\mathrm{s}^L$) and $W_\mathrm{u}^U$ ($W_\mathrm{u}^L$). 

The six panels of Figure~\ref{fig_manif_cent} are divided as follows. The upper panels refer to the high transmissivity \textcolor{black}{case}, while the lower panels correspond to the low transmissivity regime. The left (central) panels show the manifolds associated with the upper (lower) periodic orbit, while the right panels exhibit stable and unstable manifolds of different periodic orbits.

The stable and unstable manifolds associated with upper and lower orbits intersect in a complex structure, determining the motion of chaotic orbits and transport. Manifolds of the same hyperbolic orbit intersect in a structure called homoclinic tangle (see, for example, Figs.~\ref{fig_manif_cent}(b.1) and \ref{fig_manif_cent}(b.2)). \textcolor{black}{In addition, the heteroclinic tangle is defined by the intersection of manifolds of different hyperbolic orbits, see Fig.~\ref{fig_manif_cent}(a.1)}. These intersections are closely related to the emergence of chaos in Hamiltonian systems\,\cite{reichl2004}.

\textcolor{black}{Both homoclinic and heteroclinic tangles are present in the Biquadratic Nontwist Map. For example, considering the low transport configuration, Figure~\ref{fig_manif_cent}(b.1) shows the stable and unstable manifolds of the upper hyperbolic orbit in a homoclinic tangle. Meanwhile, Figure~\ref{fig_manif_cent}(b.3) shows \textcolor{black}{a} heteroclinic tangle, i.e., the intersections between the stable manifold of the upper orbit and the unstable manifold of the lower orbit. In nontwist literature, homoclinic (heteroclinic) intersections are also called intracrossing (intercrossing).}


The turnstile mechanism explains how island chains act as transport barriers based on the \textcolor{black}{homoclinic tangle} of the stable and unstable manifolds. \textcolor{black}{Fundamentally, the lobes, which are regions through which orbits enter and leave the resonance zone~\cite{meiss2015}, dictate the effectiveness of such a transport barrier}. A homoclinic (resp. heteroclininc) lobe is a region between two consecutive intersections of the stable and unstable manifolds of a given periodic orbit (resp. of a given pair of periodic orbits). In summary, the mechanism asserts that transport is directly connected to lobe size: high transport occurs with large lobe size, while low transport is associated with small lobes.

In the BNM, the homoclinic tangle differs significantly between the low and high transport configurations shown in Figure~\ref{fig_manif_cent}. The lobe sizes are considerably larger in the high transport regime [Figs.~\ref{fig_manif_cent}(a.1) and \ref{fig_manif_cent}(a.2)] compared to the low transmissivity configuration [Figs.~\ref{fig_manif_cent}(b.1) and \ref{fig_manif_cent}(b.2)]. In high transport regime, orbits can easily enter and leave the resonance zone, as stated by the turnstile mechanism. Finally, due to the symmetry \textcolor{black}{due to the symmetry between upper and lower orbits}, lobe sizes are similar, resulting in equal upward and downward transport.

\textcolor{black}{The other transport mechanism results from the structure and dominance of homoclinic and heteroclinic tangles. In low transport regimes, the homoclinic tangle dominates over the heteroclinic tangle. Comparing Figures~\ref{fig_manif_cent}(b.1) and (b.3), we note a high concentration of homoclinic intersections, in contrast to a few heteroclinic intersections. Conversely, in the high transport regime [Figs.~\ref{fig_manif_cent}(a.1) and \ref{fig_manif_cent}(a.3)], both homoclinic and heteroclinic intersections are equally distributed.}

\textcolor{black}{The above mechanism is particularly relevant in nontwist systems because they exhibit isochronous orbits whose hyperbolic manifolds can undergo a reconnection process, altering their topology. This process can change the manifold intersections from predominantly homoclinic to a denser heteroclinic tangle. Since this topological modification is a global bifurcation occurring at a critical parameter, it leads to an abrupt increase in the transmissivity of the shearless transport barrier.}

As detailed in Ref.\,\cite{szezech2009}, \textcolor{black}{heteroclinic tangles between isochronous island} chains form escape channels used by orbits to leave the transport barrier region. In the high transport regime [Figure~\ref{fig_manif_cent}(a.3)], there is a large number of \textcolor{black}{heteroclinic intersections} compared to low transmissivity, Fig.~\ref{fig_manif_cent}(b.3). We stress that both the turnstile mechanism and heteroclinic tangle are complementary in describing transport in nontwist systems\,\cite{mugnaine2018}. The turnstile mechanism dictates how orbits enter and leave the resonance zone of a specific island chain. However, since nontwist transport barriers are formed by a pair of island chains, the heteroclinic tangle governs how orbits transition between these island chains.

The manifold structure also dictates the escape channels through which orbits leave the sticky region\,\cite{portela2007}. A detailed look at Figure~\ref{fig_escTime_cent_zoom} exhibits incursions of low escape time (dark blue) among regions with significantly large escape times (light blue and green). Since the escape channels coincide with the lobes, these incursions are directly connected with the manifold behavior shown in Figure~\ref{fig_manif_cent}.

\subsection{Externally dominant}

\begin{figure}[htb]
    \centering
    \includegraphics{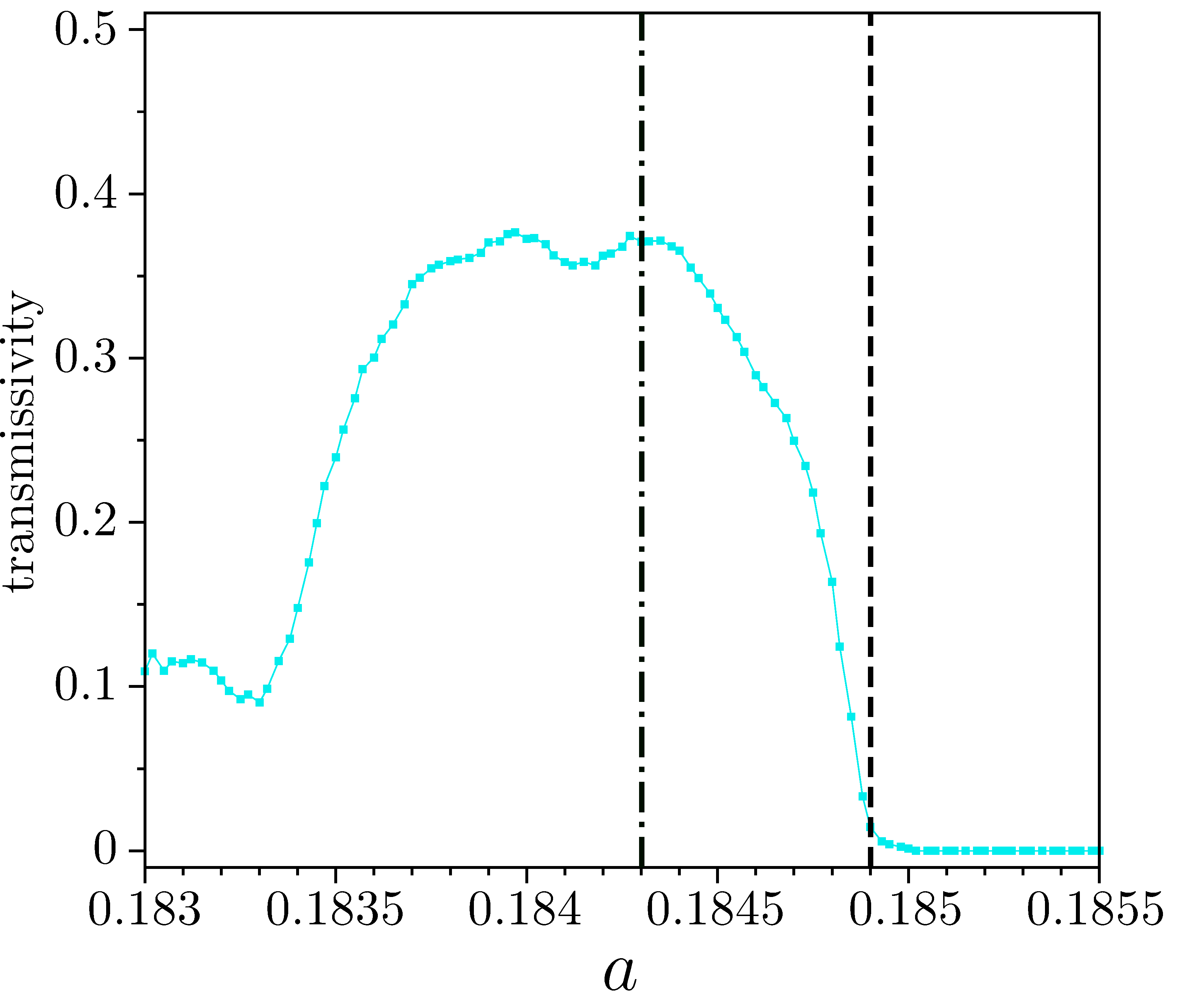}
    \caption{Transmissivity of the Biquadratic Nontwist Map, as function of $a$, fixed $b=0.77$ and $\epsilon=0.11$. Dashed-dotted and dashed lines mark high and low transport configurations, respectively. We iterated $10^5$ initial conditions up to $10^6$ times, considering boundaries at $y_\mathrm{B}=3.5$.}
    \label{fig_transm_ext}
\end{figure}

In the externally dominant scenario, processes involving the external transport barriers determine the transport properties in the Biquadratic Nontwist Map (BNM). Figure~\ref{fig_transm_ext} shows the transmissivity of the BNM as a function of the parameter $a$, considering the effect of all transport barriers combined ($y_\mathrm{B}=3.5$). Here, we observe a sudden change in the transmissivity, just like in Figure~\ref{fig_transm_central}. The configuration of low (high) transmissivity is marked by a dashed (dashed-dotted) line. As we will see, in both scenarios, the central barrier does not significantly affect the transport; only the external barriers play an important role.

\begin{figure}[htb]
    \centering
    \includegraphics[scale=1]{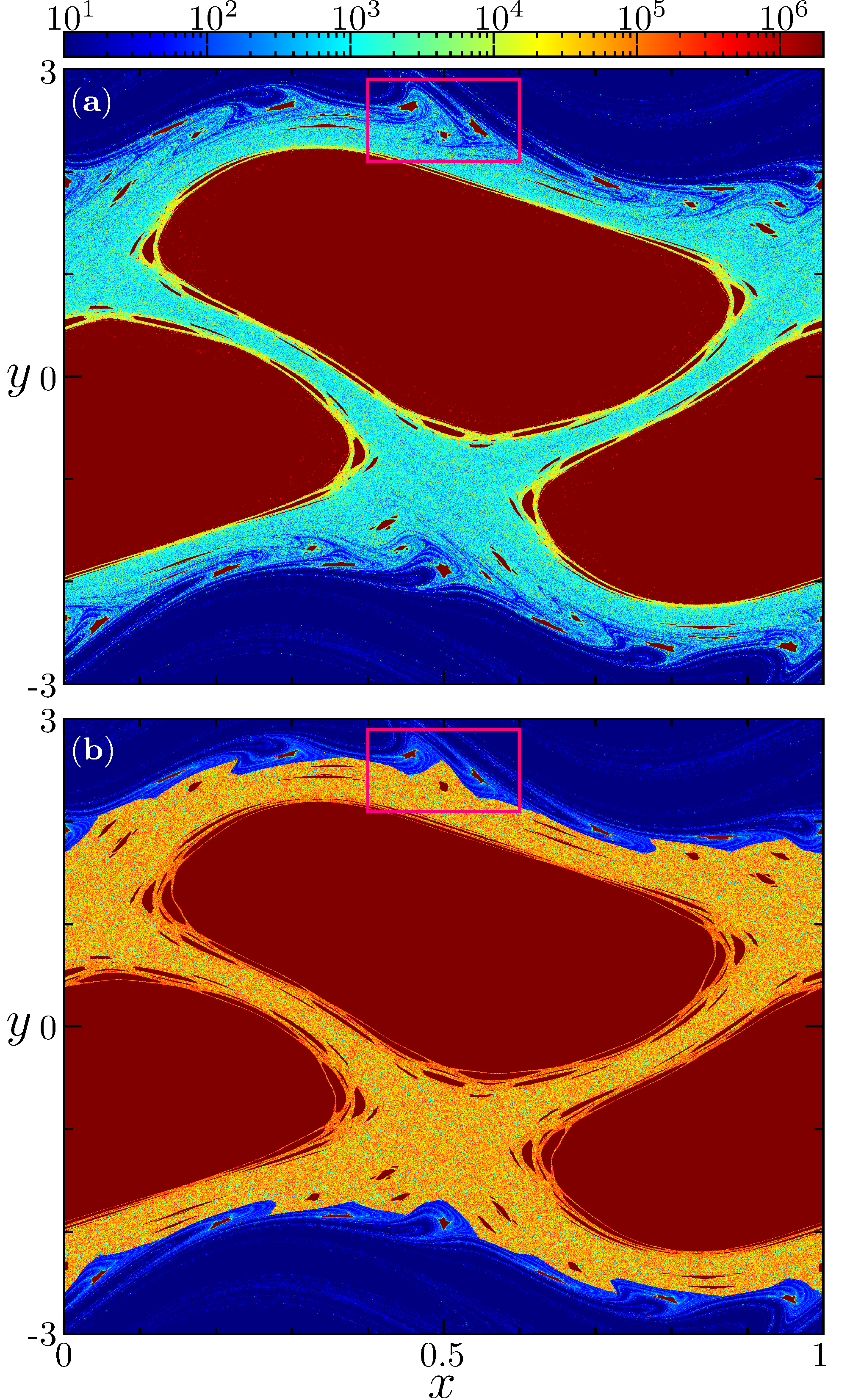}
    \caption{Escape time of the trajectories in phase space of the Biquadratic Nontwist Map, with $b=0.77$, $\epsilon=0.11$, (a) $a=0.1843$ and (b) $a=0.1849$, corresponding to high and low transport configurations of Fig.~\ref{fig_transm_ext}. Here we considered $y_\mathrm{B}=3.5$.}
    \label{fig_escTime_ext}
\end{figure}

The escape times in the high and low transmissivity configurations from Figure~\ref{fig_transm_ext} are displayed in Figures~\ref{fig_escTime_ext}(a) and ~\ref{fig_escTime_ext}(b), respectively. In both configurations, the transport barrier is characterized by an abrupt change in the average escape time, present only in the regions around $y\approx \pm 2.5$, corresponding to the external barriers. Beyond the external barriers, orbits escape after a few iterations, while between them, orbits linger to escape. This orbit trapping is more effective in the low transmissivity scenario compared to the high transmissivity one, as evidenced by the average escape time of trapped orbits.

\begin{figure}
    \centering
    \includegraphics[scale=1]{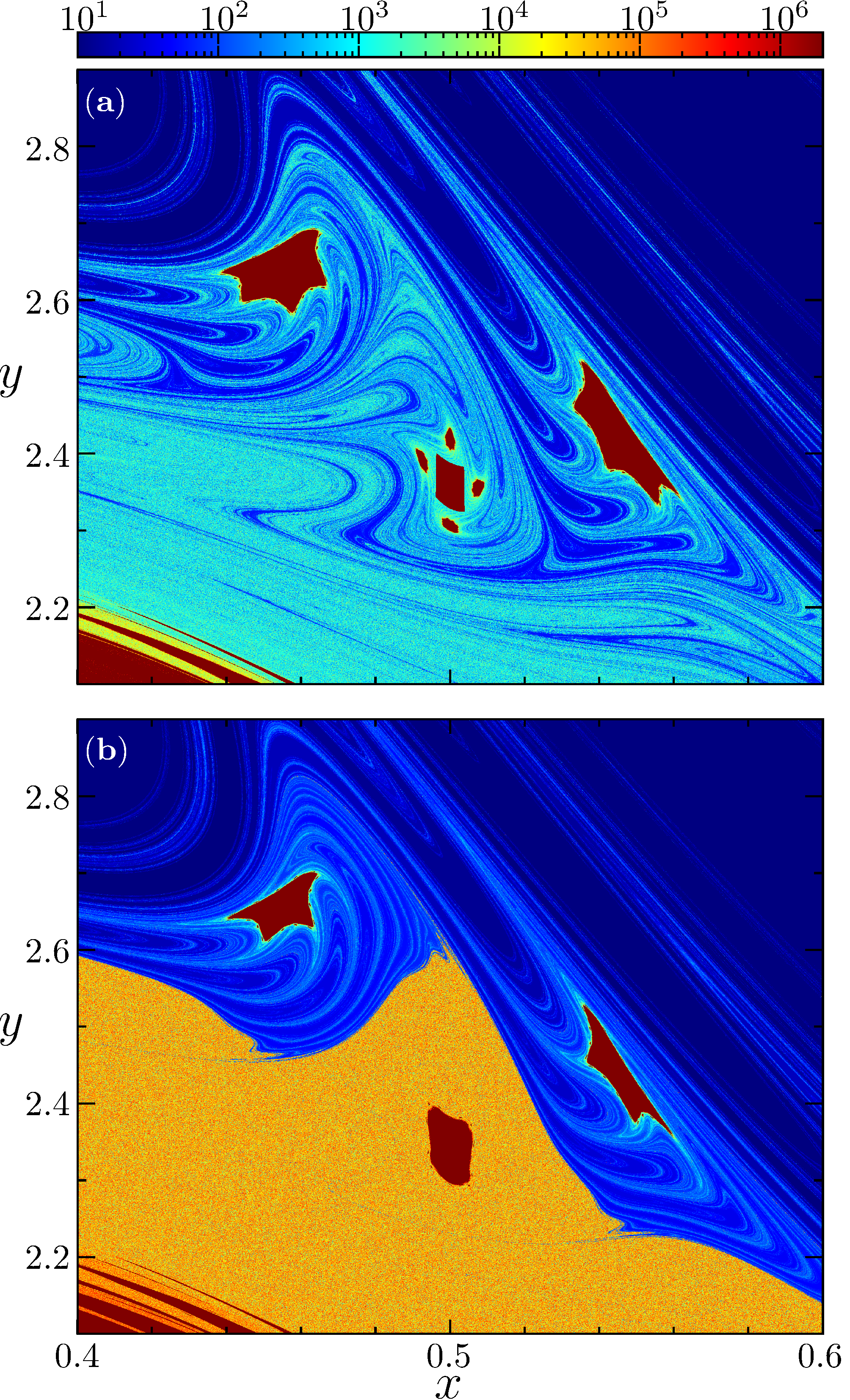}
    \caption{Magnification of the highlighted rectangle in Fig.~\ref{fig_escTime_ext}.}
    \label{fig_escTime_ext_zoom}
\end{figure}

A detailed look at the escape times near the external barriers is shown in Figure~\ref{fig_escTime_ext_zoom}, where we use the same parameters of high [Fig.~\ref{fig_escTime_ext_zoom}(a)] and low [Fig.~\ref{fig_escTime_ext_zoom}(b)] transmissivity configurations. A pair of \textcolor{black}{period-7} isochronous island chains can be seen, whose remnants are responsible for the transport barrier.

In opposition to the central transport barrier, the behavior of the map near the external barrier is asymmetric. Differences in upper and lower islands and in the average escape time are evident in Figure~\ref{fig_escTime_ext_zoom}. In both low and high transport configurations, orbits take around $10^3$ iterations to escape, except in the lower chain of low transport configuration, Figure~\ref{fig_escTime_ext_zoom}(b).

\textcolor{black}{The external shearless curves are not invariant under the symmetry transformation $S$. Consequently, although the corresponding upper and lower island chains share the same rotation number, they are not symmetric. This asymmetry generates preferred directions for chaotic transport, also known as ratchet currents. We stress the BNM itself possesses the spatial symmetry $S$, but it only guarantees equal fluxes considering symmetric boundaries, such as $\partial\mathcal{B}_\pm$. When considering the flux through only one of the external barriers, the associated boundaries are not symmetric under $S$, locally breaking the spatial symmetry. In this context, ratchet currents are possible.}

The finger-like structures in the escape time are present, dictating the escape channels of orbits. They are easily seen in Fig.~\ref{fig_escTime_ext_zoom}(a); however, in the low transport regime (see Fig.~\ref{fig_escTime_ext_zoom}(b)), they are only visible in the upper island chain due to the characteristic escape time of the region.

\begin{figure*}[htb]
    \centering
    \includegraphics[scale=1]{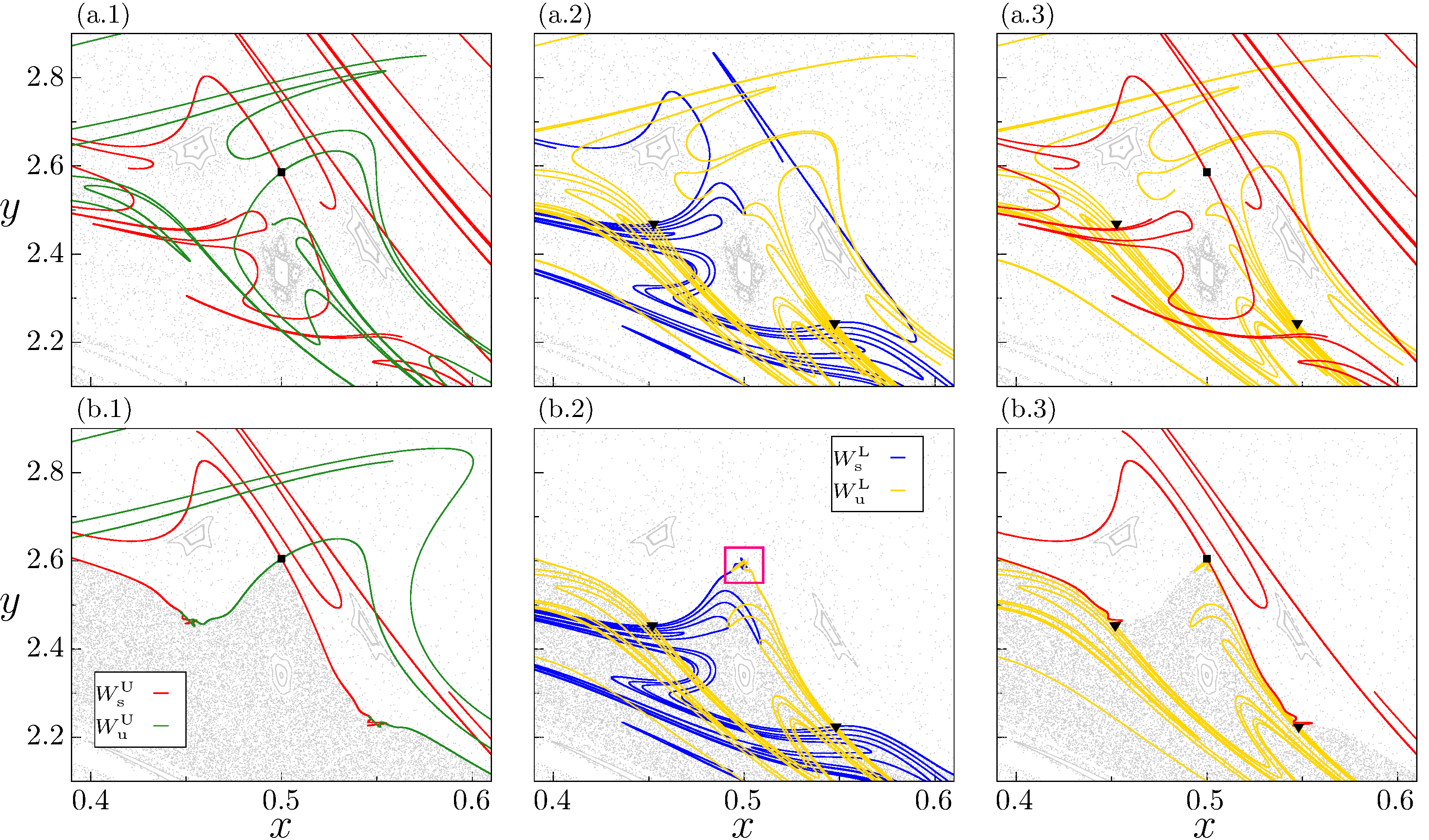}
    \caption{Stable and unstable manifolds of the upper ($W_\mathrm{s}^U$ and $W_\mathrm{u}^U$) and lower ($W_\mathrm{s}^L$ and $W_\mathrm{u}^L$) periodic orbit, considering (a) high and (b) low transport configurations of Fig.~\ref{fig_transm_ext}. Chaotic orbits near them are plotted in light-grey.}
    \label{fig_manif_ext}
\end{figure*}

The associated hyperbolic manifolds also reflect the asymmetry \textcolor{black}{between the upper and lower chains concerning the external barrier}. Figure~\ref{fig_manif_ext} shows the stable and unstable manifolds of the upper and lower island chains in Figure~\ref{fig_escTime_ext_zoom}, denoted by $W_\mathrm{s}^{U,L}$ and $W_\mathrm{u}^{U,L}$.

Following the turnstile mechanism, manifolds typically have larger \textcolor{black}{homoclinic} lobes in the high transport regime. Nevertheless, an asymmetric behavior is evident when comparing the lower and upper orbits. The lobes of the lower manifolds have roughly the same size in both high and low transport regimes, as seen in Figures~\ref{fig_manif_ext}(a.2) and \ref{fig_manif_ext}(b.2). Additionally, within the highlighted rectangle in Figure~\ref{fig_manif_ext}(b.2), small sized lobes are present. The \textcolor{black}{unequal lobe sizes indicate different} upwards and downwards transmissivity of the transport barrier.

In the externally dominant scenario, high and low transmissivity is completely determined by \textcolor{black}{the heteroclinic tangle}. Comparing Figures~\ref{fig_manif_ext}(a.3) and \ref{fig_manif_ext}(b.3), we observed a prevalence of \textcolor{black}{heteroclinic intersections} in the high transmissivity regime. \textcolor{black}{In this regime, the intersections of the hyperbolic manifolds are predominantly heteroclinic. In such cases, the upper (lower) manifolds intertwine with the lower (upper) chain, facilitating the exchange of orbits between pairs of isochronous island chains.}

Examining manifold behavior (Fig.~\ref{fig_manif_ext}) and escape time (Fig.~\ref{fig_escTime_ext_zoom}) we conclude that, in low transport regime, orbits easily enter and exit the resonance zone of the lower island chain in the external barrier. However, due to the asymmetric behavior of manifolds, the probability of these orbits finding an escape channel leading from the lower to the upper chain is low.

\section{Conclusion}
\label{sec_conclusion}

In this paper, we investigated the transport properties in the Biquadratic Nontwist Map, a prototype of a nontwist system with multiple shearless curves. Although robust to perturbations, shearless curves eventually break up; however, their remnants continue to reduce transport in the region, forming effective transport barriers. The Biquadratic Nontwist Map presents three such regions of effective barriers, referred to as the central and external transport barriers.

We used two different dynamical quantifiers to characterize the effectiveness of transport barriers: the barrier transmissivity and the escape time of orbits. The first quantifier measures the fraction of orbits that overcome the transport barrier, regardless of the time needed. The second considers the time required for each orbit to escape from the barrier region. Our results indicate that the central and external transport barriers in the Biquadratic Nontwist Map have distinct effectiveness in two identified scenarios of dominance.

In the centrally dominant scenario, the transmissivity of the central transport barrier dominates over the external barriers. In this configuration, orbits shadow the behavior of island chains, trapping them into the barrier region. Conversely, in the externally dominant scenario, the central transport barrier offers almost no resistance to transport, and the external transport barriers play a major role. In this configuration, orbits are trapped between the two external transport barriers, with escape time substantially larger than untrapped orbits.

Complementarily, we examined manifold behavior in the two dominant scenarios. As expected, the qualitative nature of manifold crossing dictates the effectiveness of the partial barriers. High transport configurations, in both scenarios, are associated with manifold crossings of different island chains (heteroclinic). However, since the map is asymmetric with respect to the external transport barriers, orbits crossing in this region have a preferred direction. This behavior is reflected in manifolds, which show varying-sized lobes. \textcolor{black}{In future investigations, a more detailed analysis of lobe sizes and their dependence on the parameters shall provide valuable insights into their influence on transport flux.} \textcolor{black}{Also, all the analyses in this work were conducted considering transport barriers associated with odd-period island chains. In the case of even-periodic chains, the scenario of individual transport barriers presents certain particularities~\cite{mugnaine2024}, which could also be explored in the context of multiple barriers.}

In summary, our results indicate that the Biquadratic Nontwist Map exhibits complex transport properties due to the presence of multiple transport barriers. Each barrier has distinct transmissivity, leading to scenarios where either the central or external barriers dominate. The behavior of manifolds, especially their \textcolor{black}{heteroclinic tangle}, plays a critical role in determining the effectiveness of these barriers. Our findings suggest that in nontwist systems with multiple transport barriers, the interplay between these barriers creates regions in phase space with significant orbit trapping, influencing overall transport dynamics.

\begin{acknowledgments}
This research was financially supported by the National Council for Scientific and Technological Development (CNPq- Grants \#309670/2023-3, \#403120/2021-7 and \#301019/2019-3) and São Paulo Research Foundation (FAPESP), Brazil. Process Numbers \#2018/03211-6, \#2022/05667-2.

The authors also thank the referees, which significantly contributed to improve our work.
\end{acknowledgments}


%

\end{document}